\documentclass[fleqn,twoside]{article}
\usepackage{espcrc2}
\usepackage{epsfig}

\usepackage{graphicx}
\usepackage[figuresright]{rotating}

\newcommand{\AmS}{{\protect\the\textfont2
  A\kern-.1667em\lower.5ex\hbox{M}\kern-.125emS}}

\title{On the scent of the {\sl knee} --- air shower measurements with KASCADE}

\author{
 J.R.~H{\"o}randel \address[1]
               {Institut f\"ur Experimentelle Kernphysik, University
              of Karlsruhe, 76021~Karlsruhe, Germany}
	      \thanks{http://www-ik.fzk.de/$\sim$joerg.},
 {T.~Antoni} \addressmark[1],
 {W.D.~Apel} \address[2]
         {Institut f\"ur Kernphysik, Forschungszentrum Karlsruhe, 
          76021~Karlsruhe, Germany},
 {F.~Badea} \address[3]
         {National Institute of Physics and Nuclear Engineering,
              7690~Bucharest, Romania},
 {K.~Bekk} \addressmark[2],
 {A.~Bercuci} \addressmark[2],
 {H.~Bl\"umer} \addressmark[1]\addressmark[2],
 {E.~Bollmann} \addressmark[2],
 {H.~Bozdog} \addressmark[3],
 {I.M.~Brancus} \addressmark[3],
 {C.~B\"uttner} \addressmark[2],
 {A.~Chilingarian} \address[4]
         {Cosmic Ray Division, Yerevan Physics Institute,
              Yerevan~36, Armenia},
 {K.~Daumiller} \addressmark[1],
 {P.~Doll} \addressmark[2],
 {J.~Engler} \addressmark[2],
 {F.~Fe{\ss}ler} \addressmark[2],
 {H.J.~Gils} \addressmark[2],
 {R.~Glasstetter} \addressmark[1],
 {R.~Haeusler} \addressmark[2],
 {A.~Haungs} \addressmark[2],
 {D.~Heck} \addressmark[2],
 {T.~Holst} \addressmark[2],
 {A.~Iwan} \addressmark[1]\address[5]
         {Department of Experimental Physics,
              University of Lodz, 90236~Lodz, Poland},
 {K-H.~Kampert} \addressmark[1]\addressmark[2],
 {H.O.~Klages} \addressmark[2],
 {J.~Knapp} \addressmark[1]\thanks{now at: 
              University of Leeds, Leeds LS2 9JT, U.K.},
 {G.~Maier} \addressmark[2],
 {H.J.~Mathes} \addressmark[2],
 {H.J.~Mayer} \addressmark[2],
 {J.~Milke} \addressmark[1],
 {M.~M\"uller} \addressmark[2],
 {R.~Obenland} \addressmark[2],
 {J.~Oehlschl\"ager}\addressmark[2],
 {M.~Petcu} \addressmark[3],
 {H.~Rebel} \addressmark[2],
 {M.~Risse} \addressmark[2],
 {M.~Roth} \addressmark[2],
 {G.~Schatz} \addressmark[2],
 {H.~Schieler} \addressmark[2],
 {J.~Scholz} \addressmark[2],
 {T.~Thouw} \addressmark[2],
 {H.~Ulrich} \addressmark[1],
 {B.~Vulpescu} \addressmark[3],
 {J.H.~Weber} \addressmark[1],
 {J.~Wentz} \addressmark[2],
 {J.~Wochele} \addressmark[2], and
 {J.~Zabierowski} \address[6]
          {Soltan Institute for Nuclear Studies, 90950~Lodz, Poland}
 }

\begin{document}

\begin{abstract}
Detailed investigations of extensive air showers have been performed with 
the data measured by
the KASCADE experiment. The results allow to evaluate hadronic interaction
models, used in simulations to interpret air shower data.
The all-particle spectrum of cosmic rays and their mass composition,
as well as individual spectra for groups of elements have been reconstructed.
The results suggest,
the {\sl knee} in the all--particle 
cosmic--ray energy spectrum is caused by a rigidity--dependent
cut--off of individual element groups.
\vspace{1pc}
\end{abstract}

\maketitle

\section{INTRODUCTION}
The earth's atmosphere is continuously bombarded by highly relativistic
ionized particles, first discovered and named "cosmic rays" by V.~Hess in 1912.
Present--day experiments show the cosmic--ray energy spectrum 
extending up to more than $10^{20}$ eV.
The flux spectrum follows a power law $dN/dE \propto E^{-\gamma}$
over many decades in energy. The most prominent feature is the {\sl knee} in
the spectrum around 3~PeV where the spectrum steepens from $\gamma\approx 2.7$
to $\gamma\approx 3.1$.
The origin of cosmic rays is still under debate. Strong, relativistic shock 
fronts expanding from supernova explosions are favoured by popular models
for the acceleration of ionized particles. Such models explain the
particle acceleration up to energies of about $Z\cdot 10^{15}$ eV, with the
nuclear charge $Z$ of the particle. This upper limit coincides for
primary protons with the
mentioned steepening of the spectrum, 
and the origin of the {\sl knee} is related
to the upper limit of acceleration in several models.

Since the charged particles are deflected in the interstellar magnetic fields,
the only hint of their sources are their energy spectrum and the mass
composition, or more preferable, the energy
spectra of individual elements. Cosmic rays at energies
below 1~PeV have been directly observed by balloon--borne instruments at the 
top of the atmosphere or in outer space. At higher energies, the steep 
falling flux 
spectrum requires
large detection areas or long observation periods, presently only possible in
ground--based installations. These detector systems measure the
secondary particles produced by cosmic rays in the atmosphere,
the extensive air showers (EAS).

To investigate the cosmic rays from several $10^{13}$ eV up to
$10^{17}$ eV the air shower experiment KASCADE ("Karlsruhe Shower Core and 
Array DEtector") \cite{kascade} has been built on--site at the
Forschungszentrum Karlsruhe in Germany.
The experiment detects the three main components of EAS simultaneously.
A $200\times 200$ m$^2$ scintillator array measures the electromagnetic and
muonic components.
The 320 m$^2$ central detector system combines a large hadron calorimeter
\cite{kalo} with several muon detection systems \cite{mwpc}.
In addition, high energetic muons are measured by an
underground muon tracking detector \cite{muontunnel}.

\section{INVESTIGATION OF HADRONIC INTERACTIONS}
In order to obtain the mass and energy of the shower--inducing particle,
measured EAS observables 
are compared with predictions of simulations. These simulations describe 
the development of an EAS in the atmosphere and, thereafter, the
signal response for all individual particles hitting the detectors.
A critical part in the simulation chain is the
model used to describe the high--energy  hadronic interactions, since
the model has to extrapolate into kinematical and energy regions not 
covered by present--day collider experiments. 
The Karlsruhe EAS simulation program 
CORSIKA \cite{corsika} provides several high--energy  hadronic interaction 
models
--- HDPM, DPMJET, NEXUS, QGSJET, SIBYLL, VENUS ---
based on different phenomenological descriptions. Below 80~GeV the
codes GHEISHA and URQMD are available.
One objective of the KASCADE experiment is to evaluate these models and to 
provide criteria for their improvement. 

\begin{figure}[hbt]
 \epsfig{file=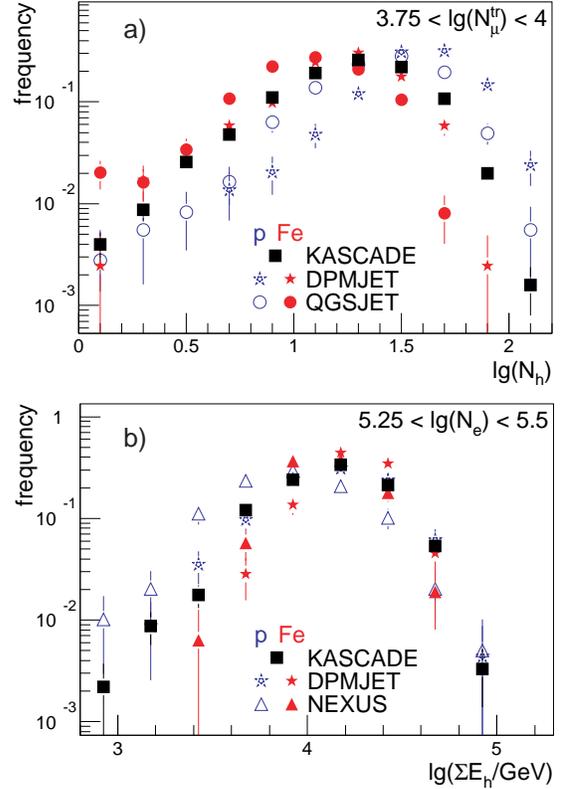,width=\columnwidth}\vspace*{-9mm}
 \caption{Frequency distribution for a) the number of reconstructed hadrons
          and b) their energy sum in EAS           
	  for two shower size intervals \cite{jensicrc}.}
 \label{jens}
\end{figure}
The hadron calorimeter is a valuable detector for testing 
the interaction models.
The structure of the hadronic component is examined in energy and 
spatial coordinates. Observables used include the number of hadrons
as well as their energy sum, their lateral distribution and their energy 
spectrum, the energy of each individual hadron relative to the most 
energetic hadron in an EAS, the maximum hadron energy, and the spatial
distribution of the hadrons. All observables are investigated
as functions of the number of electrons and muons as well as of the
hadronic energy sum. 

\begin{figure*}\vspace*{-9mm}\centering
\epsfig{file=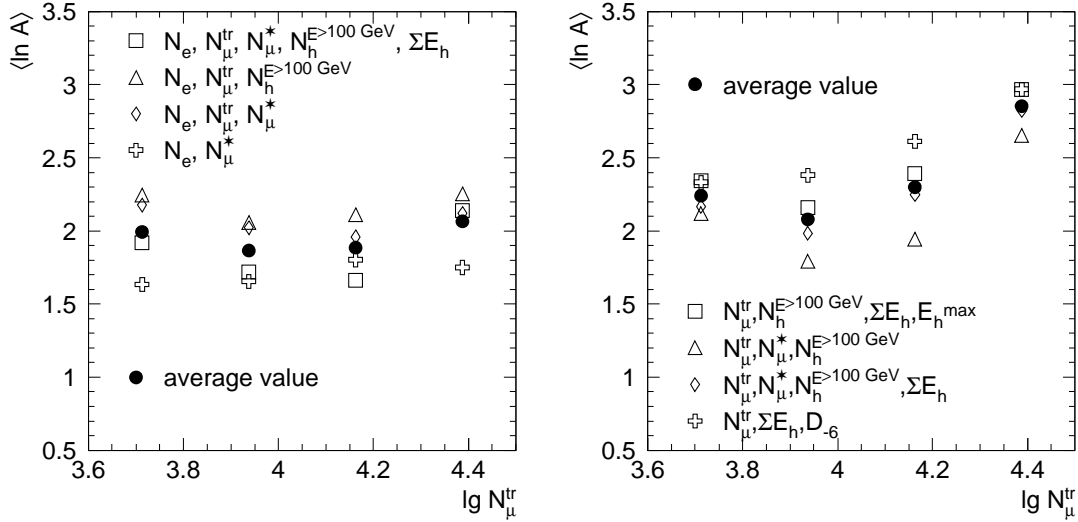,width=\textwidth}\vspace*{-12mm}
\caption{Mean logarithmic mass $\langle\ln A\rangle$
         vs. muonic shower size $N_\mu^{tr}$
         for different sets of observables \cite{roth}.}
\label{rothlna}
\end{figure*}

An example of such investigations is given in
figure \ref{jens}. Shown are frequency distributions for the number of
reconstructed hadrons above 50~GeV and the hadronic energy sum for
two shower size intervals for EAS with shower core inside the calorimeter.
KASCADE findings are compared with predictions
of the models QGSJET, DPMJET, and NEXUS for two 
extreme mass scenarios, the primaries
being only protons or iron nuclei \cite{jensicrc}.
Combining all observables, 
it turned out that the model QGSJET delivers the most 
reliable description \cite{jensicrc,wwtest}. 
This model is used for the analyses described in the following sections.

\section{ENERGY SPECTRA AND MASS COMPOSITION}
Detailed investigations have been performed with the KASCADE data
to study the influence of different observables 
and hadronic interaction models on the inferred cosmic--ray mass composition.

An example is shown in figure \ref{rothlna}.
In a Bayesian analysis \cite{roth}, the mean logarithmic mass
$\langle\ln A\rangle=\sum_i r_i A_i$, where $r_i$ is the relative abundance
of elements with mass $A_i$, is calculated for different combinations
of observables.
Several electromagnetic, muonic, and hadronic observables have been used,
including number of electrons $N_e$ and muons $N_\mu^{tr}$, number of hadrons
above 100~GeV $N_h^{100}$, their energy sum $\sum E_h$, the maximum energy of
a hadron per shower $E_h^{max}$, the number of high energy muons 
($E_\mu>2$~GeV) $N_\mu^*$, and their spatial distribution $D_{-6}$.
The left graph summarizes results obtained including the electromagnetic shower
size $N_e$, whereas the results in the right panel are obtained without $N_e$.
The combinations including $N_e$ lead to smaller $\langle\ln A\rangle$ values
as compared to those without.
The total spread between the results is 
$\Delta\langle\ln A\rangle\approx0.8$.
Similar systematic effects have been observed earlier for different
hadronic observables \cite{massealcala}.
These discrepancies are most probably caused by inconsistencies in the
hadronic interaction models used to interpret the EAS data.

Several analysis methods have been carried out to obtain the cosmic--ray
energy spectrum.
The KASCADE data show a {\sl knee} in the electromagnetic, muonic, and 
hadronic shower size spectra \cite{firstknee}.
From the size spectra, the primary energy spectrum has been derived
(see e.g. \cite{glasstettericrc}), including the absolute flux spectrum
obtained for the first time using hadronic observables \cite{hknie}.

\begin{figure}[hbt]
\epsfig{file=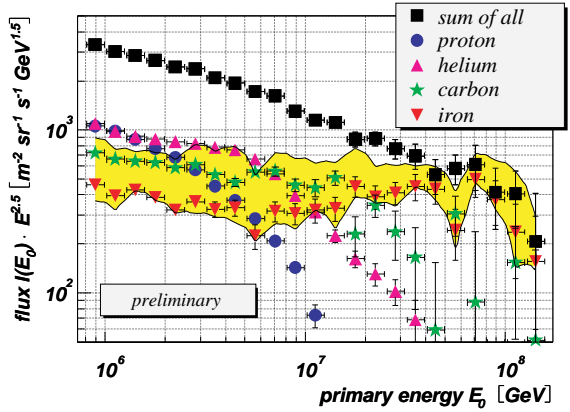,width=\columnwidth}\vspace*{-9mm}
\caption{Cosmic ray energy spectrum for four groups of elements 
         and the resulting all-particle spectrum \cite{ulrich}.}
\label{4groups}
\end{figure}

\begin{figure}[hbt]
\epsfig{file=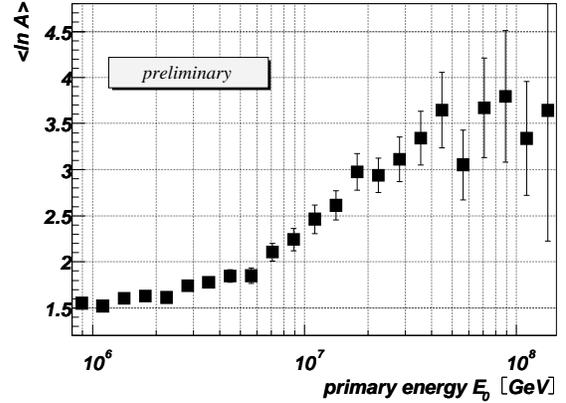,width=\columnwidth}\vspace*{-9mm}
\caption{Mean logarithmic mass vs. energy calculated from individual
 spectra for four groups of elements as shown in figure \ref{4groups}
 \cite{ulrich}.}
\label{ulrichlna}
\end{figure}

A recent analysis \cite{ulrich} uses electromagnetic and muonic shower
size spectra in three different zenith angle bins.
With a four component assumption for the mass composition of primary
cosmic rays (protons, helium, CNO-group, and iron group), an unfolding
algorithm is applied, taking into account shower fluctuations and experimental 
effects. 
The hadronic interactions have been simulated using the model QGSJET above
80~GeV and the GHEISHA code below.
Individual energy spectra for the four mass groups
are obtained as shown in figure \ref{4groups}.
Each spectrum shows a {\sl knee}--like structure.
The energy dependence of these cut--offs suggests a rigidity--dependent 
behaviour.

The mean logarithmic mass calculated from the individual energy spectra
is shown in figure \ref{ulrichlna}, indicating an increase of the
average mass above the {\sl knee}.

\section{CONCLUSION}
The EAS observables obtained by KASCADE are sensitive to hadronic interaction 
models used in simulations to interpret EAS data. At present, the combination
CORSIKA/QGSJET best describes the measurements.
The systematic dependence of the mean logarithmic mass on different
observables and models has been investigated.
The systematic error for the model QGSJET
is in the order of $\Delta\langle\ln A\rangle\approx0.8$.
The reconstruction of individual energy spectra for groups of elements
indicate a rigidity dependent cut--off, which explains the {\sl knee}
in the all particle cosmic--ray flux spectrum.

{ The KASCADE experiment is supported by the Ministry for Research of
the German government and embedded in collaborative WTZ projects
between Germany and Romania (RUM 97/014), Poland (POL 99/005), and
Armenia (ARM 98/002). The Polish group acknowledges the support
by KBN grant no. 5\,P03B\,133\,20.
}

\end{document}